\documentclass[a4paper]{article}

\usepackage{INTERSPEECH2022}
\usepackage{cite}

\usepackage{color}

\title{Leveraging Phone Mask Training for Phonetic-Reduction-Robust E2E Uyghur Speech Recognition}
\name{Guodong Ma$^1$, Pengfei Hu$^2$, Jian Kang$^2$, Shen Huang$^{2*}$, Hao Huang$^{1*}$
\thanks{$^*$ corresponding authors}}
\address{
  $^1$School of Information Science and Engineering, Xinjiang University, Urumqi, China \\
  $^2$Tencent Minority-Mandarin Translation, Beijing, China}
\email{\{mag431217, hwanghao\}@gmail.com, studenthu@163.com, kangjianqh@sina.com, qqhuangshen@hotmail.com}

\begin{document}

\maketitle
\begin{abstract}
In Uyghur speech, consonant and vowel reduction are often encountered, especially in spontaneous speech with high speech rate, which will cause a degradation of speech recognition performance. To solve this problem, we propose an effective phone mask training method for Conformer-based Uyghur end-to-end (E2E) speech recognition. The idea is to randomly mask off a certain percentage features of phones during model training, which simulates the above verbal phenomena and facilitates E2E model to learn more contextual information. According to experiments, the above issues can be greatly alleviated. In addition, deep investigations are carried out into different units in masking, which shows the effectiveness of our proposed masking unit. We also further study the masking method and optimize filling strategy of phone mask. Finally, compared with Conformer-based E2E baseline without mask training, our model demonstrates about 5.51\% relative Word Error Rate (WER) reduction on reading speech and 12.92\% on spontaneous speech, respectively. The above approach has also been verified on test-set of open-source data THUYG-20, which shows 20\% relative improvements.

\end{abstract}


\section{Introduction}
In recent years, with the rapid development of neural network, the performance of ASR system has been greatly improved. Current traditional ASR systems consists of several components including acoustic, lexicon, and language models. E2E ASR systems train these models jointly and the systems map acoustic feature sequences to token sequences directly. E2E ASR framework can generally be divided into the following categories: connectionist temporal classification (CTC) \cite{ctc-paper,RNNt-paper} and attention-based sequence-to-sequence encoder-decoder \cite{LAS-paper,Speech-Transformer,Conformer}. In this context, many works are based on E2E frameworks to explore different ASR scenarios and improve the performance accordingly. For example, code-switch \cite{Code-switch1, Code-swith2, Code-swith3, code-switch4} and low-resource task \cite{Low-Resourced1, Low-Resourced2, Low-Resourced3} etc.

The Uyghur language, which belongs to Altaic language family, is mainly spoken by a large population in Xinjiang, China. Research on Uyghur ASR has been carried out for many years. The study on ASR of the Altaic language family generally focuses on issues, such as agglutinative characteristics and low resources \cite{Uygur-paper1, Uygur-paper2, Uygur-paper3, Uygur-paper4, Uygur-paper5, Uygur-paper6, Uygur-paper7}.
Altaic languages, such as Uyghur, is spoken much faster than other languages, such as Mandarin, English, etc. Therefore, in the daily life of Uyghur, there will be common language phenomena such as consonant and vowel reduction, which is referred as phonetic reduction (PR). The phones which have undergone PR become shorter. 
Accordingly, it brings great difficulties to the construction of robust Uyghur ASR. 
In the conventional ASR, PR phenomena can be straightforwardly modeled by using a multi-pronunciation dictionary, in which the possible PR pronunciation variants are included. 
But it is difficult to get all possible pronunciation variants normally, and need require the knowledge of language experts. In E2E ASR systems, the learning of pronunciation dictionary information is integrated in the model, so how to modify the model framework to learn dictionary information is a key question to explore. 

We try to solve PR from a mask training perspective under the E2E ASR framework.
Recently, a lot of excellent works about masking have been proposed. BERT \cite{BERT} masks the modeling units so that the model can consider more information of the context. SpecAugment \cite{SpecAugment} masks the spectrum of time or frequency domain in random. It increases the diversity of training data to prevent over-fitting. 
 In addition, semantic mask \cite{SemanticMask} randomly masks spectrum of word of 15\% during training E2E model. It aims to improve the strength of LM. However, the lack acoustic meaning of masking, the random and large masking size, make above algorithms cannot be connected with the problems we need to solve directly.
Inspired by the above masking methods \cite{BERT, SpecAugment, SemanticMask} and the peculiarities in Uyghur, we propose an effective masking method, phone mask (PM).  The main idea is to randomly mask off a certain percentage features of phones during model training. Having such case in Uyghur of which a phone stands for a grapheme (except individual phone), so phone mask generally equal to grapheme mask. It simulates the above verbal phenomena and facilities model to learn more contextual information. In addition, we further to discuss the masking ways and filling strategies of phone mask. Seeing section 3 for details. Compared with  E2E model baseline without masking training, our model indicates an WER relative reduction of 5.51\% on reading speech and 12.92\% on spontaneous.

Our main contributions are as follows: To our best knowledge, we are the first to use a large amount of Uyghur speech data to explore ASR task under the  E2E framework. And also perhaps we are the first to address the issue of PR and provide a solution to this problem under the Conformer based E2E ASR framework. Moreover, through experiments and investigations into a variety of masking configurations are presented. 


\section{Related Work}
Recent research on ASR of Altaic languages such as Uyghur are mainly focusing on the solution to the OOVs problem caused by the stickiness and under-resourced problem caused by data scarcity \cite{Uygur-paper1, Uygur-paper2,Uygur-paper3, Uygur-paper4, Uygur-paper5, Uygur-paper6, Uygur-paper7}. To our best knowledge, there were no relevant papers that had researched the solutions to the problems of PR for ASR. Recently, emerging masking methods such as BERT \cite{BERT}, SpecAugment \cite{SpecAugment} and semantic mask \cite{SemanticMask} has shown to be very successful. SpecAugment \cite{SpecAugment} focuses more on the diversity of data onto model training, so as to prevent to over-fitting. The other two masking strategies \cite{BERT, SemanticMask} concern more about utilizing the surrounding information as much as possible to learn  abundant LM knowledge and improve the robustness of model. The PR occurrence depends greatly on the contextual pronunciations and might be learned through large amount speech data. Therefore, application of the mask training to solving the PR problem in Altaic language, such as Uyghur, can be reasonable and feasible. 

\section{Proposed Method}

\subsection{ Phonetic reduction in Uyghur speech}

Phonetic reduction are very common in Altaic languages \footnote{https://en.wikipedia.org/wiki/Altaic\_languages} \footnote{https://en.wikipedia.org/wiki/Uyghur\_language}, especially in high speed spontaneous speech. To give a clearly illustration, we built a conventional hybrid ASR model using  Kaldi \cite{Kaldi} and compute length of  each phone by doing phone-level forced alignment. 
Firstly we compare the average length of phonemes in the languages of Mandarin, English and Uyghur. As shown in Table~\ref{data:phone_duration}, the average duration of Uyghur phones are obviously shorter than that of Mandarin and English, while Mandarin has the longest phone duration.

According to analysis of the forced alignment results, Uyghur has higher ratios of phones which duration is 3 frames. Since the HMM models used in force alignment has 3 states,  every phone has 3 frames duration at least during force alignment.  We check these short duration phones' speech and find that many of them has phonetic reduction phenomenon to some extent. It is also found that the high speaking speed aggravate phonetic reduction in Uyghur.


\begin{table}[htbp]
  \caption{The average duration of phonemes}
  \label{data:phone_duration}
  \centering
  \begin{tabular}{l|lll}
    {Language} & {Duration (Seconds)}    \\
    \midrule
    \hfil Mandarin          & \hfil 0.14            \\
     \hfil English           & \hfil 0.10            \\
    \hfil Uyghur            & \hfil  0.08             \\
  \end{tabular}
\end{table}

\subsection{Masking Training}

\begin{figure}[t]
  \centering
  \includegraphics[width=\linewidth]{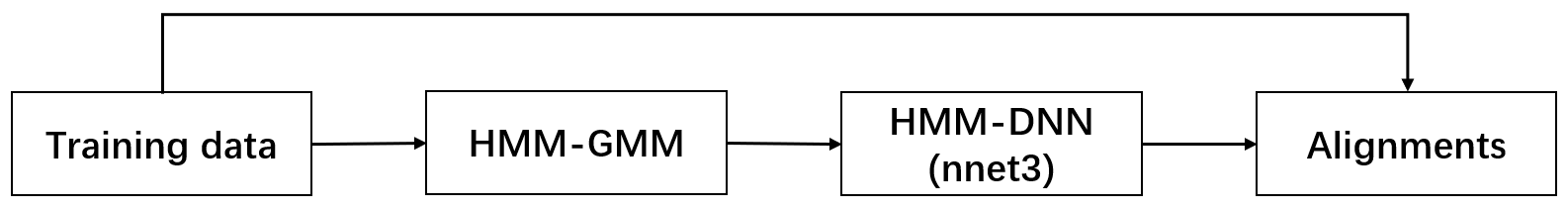}
  \caption{Get alignments using kaldi}
  \label{fig:get_alignments}
\end{figure}

\begin{figure}[t]
  \centering
  \includegraphics[width=\linewidth]{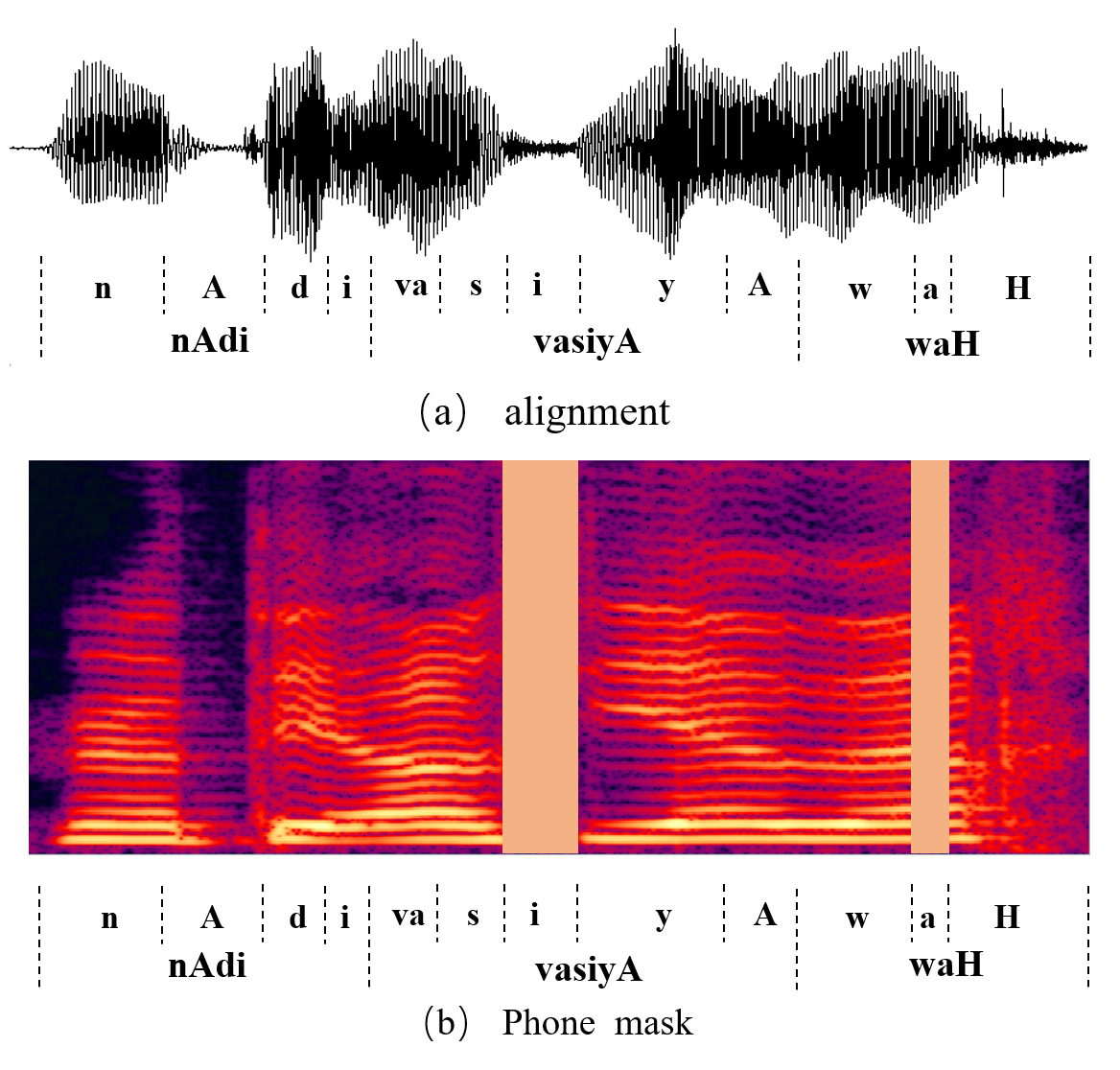}
  \caption{Example of alignment and phone mask}
  \label{fig:example_mask}
\end{figure}

The masking training of E2E speech recognition can be traced back to SpecAugment \cite{SpecAugment}, which improved the system robustness by masking spectrum randomly in time or frequency domain. Instead of randomly masking, the semantic mask method masks the acoustic features corresponding to a particular word, which improves the power of the implicit language model in the decoder. Aiming at reducing the impact of  phonetic reduction in Uyghur speech,  we propose the phone mask method, which masks the acoustic features of particular phone.  To achieve this, we need to obtain alignment information on the training data. As shown in  Figure~\ref{fig:get_alignments}, we train an HMM-DNN (nnet3) model using Kaldi and get the alignment information of each phone and frame using forced-alignment. During model training, we randomly select a percentage of the phones and mask the corresponding speech segments in each iteration as shown in Figure~\ref{fig:example_mask}. In experiment part later, we make a deep study about masking ratio and the filling strategy. 

The  proposed phone mask training is to some extent similar to semantic mask. They both mask spectrum during model training. But the biggest difference between our method and semantic mask \cite{SemanticMask} is that our method not only to encourage E2E model to learn more contextual information, but also to make it to learn more dictionary knowledge.  The intuitions behind these  methods are also different. Our proposed phone mask aims to phonetic reduction characteristic of the Uyghur speech. Meanwhile, we think that the word-level semantic masking has too large masking span and will put great pressure for the model learning. We will experiment with different masking granularity later to prove this conjecture. 

\section{Model}

\subsection{Conformer Encoder}
Conformer \cite{Conformer} is a SOTA ASR encoder architecture, which inserts a convolution layer into transformer block to increase the local information modeling capability of the traditional Transformer model. The architecture of Conformer 
encoder consists of a self-attention, convolution, feed-forward, and layernorm modules. The second feed-forward module is followed by a final layernorm layer. For example $x$ is the input of the Conformer block $i$, then the $i$-th block output is calculated as
\begin{eqnarray}
    x' &=& x + \frac{1}{2}FFN(x)\\
    x'' &=& x' + MHSA(x')\\
    x''' &=& x'' + Conv(x'')\\
    y_i &=& Layernorm(x''' + \frac{1}{2}FFN(x'''))
\end{eqnarray}
where FFN(·), MHSA(·) and Conv(·) denote the Feed-Forward Network Module, Multi-Head Self-Attention Module and Convolution Module.

\subsection{ASR Training and Decoding}
Following \cite{CTC_conformer, recent-conformer}, we use Conformer (CE) and CTC objectives for training and decoding. In the training step, all models are trained to minimize both Conformer (CE) and CTC losses jointly. The objective function is described as follows:

\begin{equation}
    Loss = - \alpha Loss_{ce} - (1 - \alpha) Loss_{ctc}
\end{equation}
where $\alpha$, a tunable parameter, denotes an weight that balances the Conformer loss (CE loss) and CTC loss. In this paper, we set $\alpha$ = 0.7. In the decoding stage, we combine the probabilities of Conformer with CTC as follows:
\begin{equation}
    \hat{Y} = \arg\underset{y}{\max} (\lambda\log P_{ce}(y|X) + (1-\lambda)\log P_{ctc}(y|X))
\end{equation}
where $X$ represents the inputs and $\lambda$ is a tunable parameter. Following \cite{SemanticMask}, we set $\lambda$  to 0.5.

\section{Experiments and Results}

\subsection{Experimental setup}

For the training data, we collect 1200 hours Uyghur speech corpus from several data companies. As for the evaluation set, we use speech datasets collected from two scenarios, one is reading speech (Read-Test) and the other is spontaneous speech (Oral-Test). Among them, the speech rate of Oral-Test set is much faster than the Read-Test set. The details about the datasets are shown in Table~\ref{data:details}. All the experiments use 40 Mel Frequency Cepstral Coefficients (MFCCs) over 25 ms frames with 10 ms stride to each of which cepstral mean and variance normalization (CMVN) is applied. We both experiment using the state-of-the-art hybrid and E2E ASR systems. We add 100 i-Vector only to the hybrid ASR system for speaker adaptation. 


\begin{table}[htbp]
  \caption{Descriptions of the datasets }
  \label{data:details}
  \centering
  \begin{tabular}{l|cccc}
    & {Duration}    &  & Avg Phone\\
        {Data Type}  & {(Hours)}    & Domain & Dur. (Sec.) \\
    \midrule
    Train & 1200 & Reading/Clean & 0.09  \\
    Read-Test & 3& Reading/Clean &0.10  \\
    Oral-Test & 5& Oral/Noisy & 0.06     \\
  \end{tabular}
\end{table}

For the hybrid ASR system, we adopt the chain model \cite{chain-model-papaer} using Kaldi toolkit \cite{Kaldi}. The neural network has 15 hidden layer TDNN and a rank reduction layer. The number of units in TDNN consists of 1563 and 160 bottleneck unit. The learning rate for training with 5 epochs is chosen initially from 0.00025 ending at 0.000025. We have developed 4-gram LM using the transcription of training data trained on SRILM toolkit \cite{SRILM-work}. 

As for the E2E system, we adopt the Conformer \cite{Conformer} and Transformer architecture of \cite{SemanticMask}, named STM-Transformer. For comparability, the configurations of our Conformer are the same as STM-Transformer (Encoder = 12, Decoder = 6, H = 8, $d^{att}$= 512). Other configurations of Conformer follow \cite{recent-conformer}. The output tokens are 5000 BPE tokens produced by unigram model using sentencepiece \cite{SentencePiece}. All the E2E models are trained by using the speech recognition toolkit ESPnet \cite{espnet} on 4 P40 GPUs, which are trained 40 epochs (decreased epochs due to speed perturbation \cite{speed-perturb}) spending about 5 days.

\subsection{Results}

\subsubsection{Results on Conformer}

To our best knowledge, no related work before explored the E2E Uyghur speech recognition using a large amount of speech data, so we firstly conduct the experiments on grapheme-based and word-piece-based Conformer and compare the results with the chain model.  It can be seen that the Conformer based system show better performance than the chain model on the Oral-Test, but the grapheme-based Conformer show slightly worse performance than the chain model on the Read-Test. The reason might be that the transcripts used in LM of chain model and Read-Test set belong to similar domains, which enhances the performance of Read-Test with Chain model, while grapheme-based Conformer's decoder is hard to model the relationship across words.  Similar to general E2E ASR, word-piece-based Conformer outperforms chain model and grapheme-based Conformer on both of the two test sets.


\begin{table}[t]
  \caption{Comparisons between semantic mask (STM), word piece mask(WPM), phone mask (PM), and PM variants using Conformer\cite{Conformer}; Mask methods\_(number) mean the mask ratio is number, if have no number, 15 \% by default. (WER)}
  \label{phone:mask}
  \centering
  \begin{tabular}{l|cccc}
    {SYSTEM} & {Read-Test}  & {Oral-Test}   \\
    \midrule
    Chain                   & 26.0           &47.9           \\
    Grapheme-based Conformer      & 26.8            &45.0            \\
    \hline
    Word-piece-based Conformer    & 25.4            &44.1            \\
    \quad + SpecAugment     & 25.2            &41.6            \\
    \quad + STM             & 25.1            &42.4            \\
    \quad + WPM             & 25.0            &40.1            \\
    \quad + PM              & 24.8            &38.9            \\    
    \quad + WPM\_20         & 24.4            &40.0            \\

    \quad + PM\_20          & 24.2            &38.9            \\
    \quad + PM\_20 + \_FW     & \textbf{24.0}   &\textbf{38.4}   \\
  \end{tabular}
\end{table}


For mask training, we firstly experiment on different masking methods including SpecAugment, semantic mask (STM), and phone mask (PM).
By comparing the results of SpecAugment with STM, 
it can be observed that the former performs better on the Oral-Test set. We think it is because the word-level STM has a relative larger masking granularity, which put great overhead on model learning. However, SpecAugment, especially time masking, is capable of alleviating the consonant and vowel reduction in Uyghur to some extent. Overall, our proposed PM, which is inspired by the peculiarities of Uyghur, achieves the best performance among them, which show the effectiveness for phonetic reduction and further demonstrate the granularity play an important role in masking.
Secondly, in order to more effectively illustrate the issue of the large masking granularity of STM, we replace the word-level masking (STM)  with word-piece-level, which is named word-piece mask (WPM). The results by WPM show  better than those from word-level masking (STM) and SpecAugment. This is consistent with our hypothesis that it hurt the WER if the masking granularity is too large.  


We also experiment with different mask ratio, i.e. the percentage of phone or word-piece that are masked in the training process. The default mask ratio is set to 15\% in the above experiments, following the BERT \cite{BERT} and semantic mask \cite{SemanticMask}. We further adjust the ratio of masking from 15\% to 30\% for the experiment. We have observed the masking ratio of 20\% is the optimum for phone mask and word piece mask. The results are also shown in Table~\ref{phone:mask}, which also indicates that the granularity of word-level masking affects the effect of masking to some extent. As for filling strategy (PM\_20 + \_FW), besides filling with the average value of the whole utterance features, we also try to use the average value of word where the selected masked phone is located. It enables the model to make use of more relevant information. We can see the performance, in the last row of Table~\ref{phone:mask}, has been further improved.



As mentioned in Section 3, phonetic reduction can be aggravated by high speaking speed. Therefore, it is  naturally to consider the utilization of speed perturbation \cite{speed-perturb} to augment the speech data of variant speed. In Table~\ref{phone:speed_perturbation}, we experiment on speed perturbation, PM, and PM after speed perturbation. By comparing  speed perturbation only (SP) and PM only, we can see that pure PM achieves better performance than SP on both of the two test sets. Especially, the improvements on the Oral-Test set is more obvious, which further demonstrates the effectiveness of the proposed PM training. Finally, when we do PM after SP, it shows a 2.9\% absolute improvement on Oral-Test and 1.0\% on Read-Test compared with SP only system.


\begin{table}[htbp]
  \caption{Comparison with speed perturbation (WER)}
  \label{phone:speed_perturbation}
  \centering
  \begin{tabular}{l|ccc}
    {SYSTEM} & {Read-Test}  & {Oral-Test}   \\
    \midrule
    Word-piece-based Conformer   & 25.4            &44.1            \\
    \quad + PM\_20 + \_FW        &24.0             &38.4            \\
    \quad + Speed Perturb        & 24.5            &39.9            \\
    \quad\quad + PM\_20 + \_FW   & \textbf{23.5}   &\textbf{37.0}   \\
  \end{tabular}
\end{table}



\subsubsection{Results on the THUYG-20 benchmark's test set}

\begin{figure}[t]
  \centering
  \includegraphics[width=8.2cm]{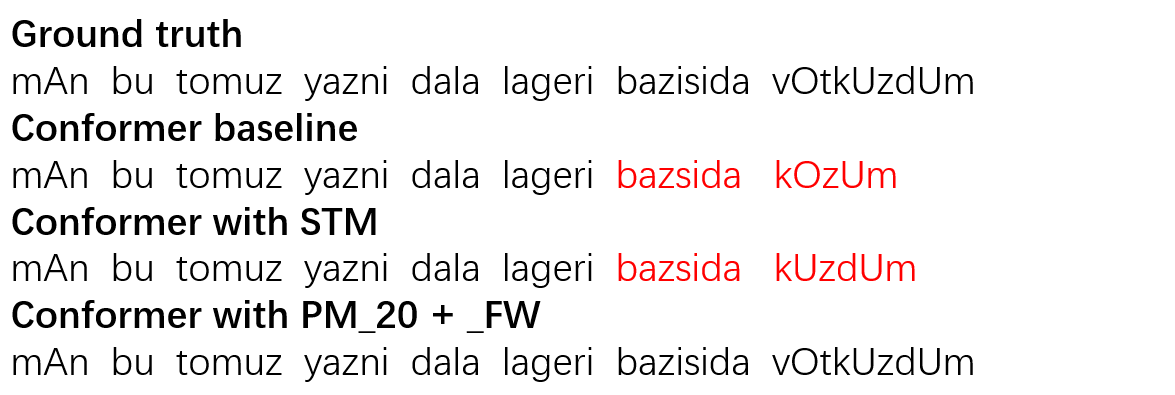}
  \caption{Case  of  THUYG-20 test-set (UTT\_ID: F1010\_037)}
  \label{Test:Thuyg-case}
\end{figure}

To further demonstrate the effectiveness of the proposed method, we present the results on a popular open-source Uyghur ASR benchmark, THUYG-20  \cite{Thuyg-20}. Table~\ref{Test:Thuyg-test} shows the results. The results are in consistence with the conclusions reached above. The PM\_20 + \_FW system shows the best performance with a WER of 10.0\% on the THUYG-20 test set. To better illustrate how the proposed method alleviate the problem of phonetic reduction, in Figure~\ref{Test:Thuyg-case}, we show a case of a THUYG-20 test utterance with its ground-truth  label and the corresponding recognition outputs using different model configurations. 
We can see the issue of phonetic reduction can be solved by the proposed method from the perspective of masking. 
The results show that the phone mask we proposed can better reduce substitution errors and achieves the best WER.

\begin{table}[htbp]
  \caption{WER   on the THUYG-20 test set }
  \label{Test:Thuyg-test}
  \centering
  \begin{tabular}{l|lllll}
    {SYSTEM} & {WER}    & SUB & DEL & INS \\
    \midrule
    Base Conformer                & 12.5           & 10.4 & 1.4 & 0.7             \\
    \quad\quad+ SpecAugment       & 12.6           &  10.3 & 1.4 & 0.9             \\
    \quad\quad+ STM               & 12.4           & 10.1 &1.5 &0.8             \\
    \quad\quad+ PM\_20            & 10.6           & 8.7 & 1.4 & 0.5             \\
    \quad\quad+ PM\_20 + \_FW     & \textbf{10.0}  & \textbf{8.3} &\textbf{1.2} &\textbf{0.5}             \\
  \end{tabular}
\end{table}

%

\subsubsection{Results on STM-Transformer}
To provide a fair comparison and the applicability  of our proposed method to other network structure, we make an experimental comparison on  STM, WPM and our proposed PM using the STM-Transformer network structure and configuration proposed on \cite{SemanticMask}. Table~\ref{semantic:architecture} presents the results (WER). Similar to \cite{SemanticMask}, the baseline Transformer represents the model with position embedding.  The proposed PM training  still achieves the best performance on the two test sets. Compared with the baseline and STM, the PM\_20 model show absolute improvements of 5.2\% and 3.3\% on the Oral-Test set, respectively, and a significant improvement on the Read-Test set as well. The WER reduction on the Oral-Test set is greater than that  on the Read-Test set. Meanwhile, we can see from the Table~\ref{semantic:architecture} that, both Read-Test and Oral-Test, WPM is better than STM, which further confirms our guess in Section 3 that too large masking span  will put great overhead for the model learning.

\begin{table}[htbp]
  \caption{Comparison of mask training using STM-Transformer}
  \label{semantic:architecture}
  \centering
  \begin{tabular}{l|cc}
    {SYSTEM} & {Read-Test}    & Oral-Test   \\
    \midrule
    Base Transformer                       & \hfil 28.9              &\hfil 44.7              \\
    STM-Transformer                        & \hfil 27.9              &\hfil 44.3              \\
    STM-Transformer with STM               & \hfil 25.6              &\hfil 42.8              \\
    STM-Transformer with WPM\_20           & \hfil 25.1              &\hfil 41.7              \\
    STM-Transformer with  PM\_20           & \hfil \textbf{24.5}     &\hfil \textbf{39.5}     \\
  \end{tabular}
\end{table}



\section{Conclusions}

In this paper, aiming at resolving the phenomenon of phonetic reduction in Altaic language, such as Uyghur, we propose an effective phone masking method for E2E speech recognition. During training, a certain proportion spectrum of phones will be masked and filled with the average of word span spectrum features. Therefore, more contextual information and dictionary knowledge are expected to be learned with the help of phone masking. We have carried out Uyghur speech recognition tests on both reading and spontaneous speech. Results show that proposed method improves the performance of Uyghur speech recognition obviously, especially on the spontaneous speech. Extensive investigations into the masking granularity, comparison with traditional masking methods and applicability to other model structure have also been carried out and confirm the effectiveness of the proposed method.  According to the analysis, the phone masking training is very helpful to improve the robustness of Uyghur speech recognition system to PR problem.

\section{Acknowledgements}

The authors of this paper would like to give special thanks to Dr. Nurmemet Yolwas with Xinjiang University for guidance and help in this work. This work was supported by the National Key R\&D Program of China (2017YFB1402101) and Natural Science Foundation of China (61663044, 61761041). 



\vfill\pagebreak

\bibliographystyle{IEEEtran}

\bibliography{mybib}


\end{document}